\documentclass[pra,showpacs,twocolumn,showkeys]{revtex4-1}
\usepackage{graphics}
\usepackage{amsmath}
\usepackage{slashed}
\usepackage{mathtools}
\usepackage{color}
\usepackage{bbold}
\usepackage[usenames,dvipsnames]{xcolor}
\usepackage{dsfont}
\usepackage{braket}
\begin{document}
\newcommand{\bstfile}{osa}
\newcommand{\bibs}{d:/Dropbox/Dad/Mark/References/BibFile}
\title{Quantum Origins of the Density Operator}
\author{Mark G. Kuzyk}
\affiliation{Department of Physics and Astronomy, Washington State University, Pullman, Washington  99164-2814
\\ \today}
\begin{abstract}
Students in quantum mechanics class are taught that the wave function contains all knowable information about an isolated system.  Later in the course, this view seems to be contradicted by the mysterious density matrix, which introduces a new set of probabilities in addition to those that are built into the wave function.  This paper brings attention to the fact that the density matrix can be reconciled with the underlying quantum-mechanical description using a two-particle entangled state with a one-particle subsystem as the simplest illustration of the basic principle.  The extra-quantum probabilities are traced to the coefficients of superposition of the quantum state vector and the seemingly irreversible exponential population decay is shown to be compatible with the unitary time evolution of a pure state when the two particles interact.  The two-particle universe thus provides the student with a tool for understanding how the density operator, with all its richness, emerges from quantum mechanics.
\end{abstract}

\maketitle
\tableofcontents
\section{Introduction}\label{sec:intro}

The textbook development\cite{sakur10.01,sakur94.01,huang63.01,pathr01.11,gottf01.74} and pedagogical treatment\cite{patte62.01,sabba07.01} of the density operator\cite{neuma27.01,neuma18.01} typically calls upon a hybrid classical/quantum argument as follows.  Consider a collection of particles, such as electrons.  If $w_i$ represents the fraction of electrons in state $i$, the expectation value of the observable represented by the operator $A$ is given by
\begin{equation}\label{Eq:ExpectA}
\Braket{A} = \sum_{i} w_i \Braket{i|A|i} .
\end{equation}
Equation \ref{Eq:ExpectA} makes intuitive sense; $\Braket{i|A|i}$ is the {\em quantum} expectation value of observable $A$ in state $i$ and $w_i$ provides the {\em statistical weighting} to the quantum expectations.

The index $i$ can span a number of states that exceeds the dimensionality of the Hilbert space of the quantum system and the individual states in the sum need not be orthogonal, so it is often the case that $\Braket{i | i^\prime} \neq \delta_{i,i^\prime}$.  For example, the spin (living in a two-dimensional space) can have a fifth of the electrons oriented along $\hat{z}$, a tenth along $\hat{x}$, a quarter along $(\hat{x} + \hat{y})/\sqrt{2}$, and so on.  As such, closure need not apply to $\Ket{i}$, that is $ \sum_i \Ket{i} \Bra{i} \neq \mathds{1}$.

For a complete orthonormal basis $\lbrace \Ket{n} \rbrace$, closure holds and can be inserted into Equation \ref{Eq:ExpectA}, yielding
\begin{equation}\label{Eq:ExpectAclose}
\Braket{A} = \sum_n \sum_{i} w_i \Braket{i| (\Ket{n} \Bra{n}) A|i} .
\end{equation}
Interchanging the scalars $\Braket{n|A|i}$ and $\Braket{i|n}$, Equation \ref{Eq:ExpectAclose} takes the form
\begin{equation}\label{Eq:ExpectAinterchange}
\Braket{A} = \sum_n \sum_{i} w_i \Braket{n|A|i} \Braket{i|n}.
\end{equation}
Defining the operator
\begin{equation}\label{Eq:DensityOp}
\rho = \sum_i w_i \Ket{i} \Bra{i},
\end{equation}
Equation \ref{Eq:ExpectAinterchange} can be expressed as
\begin{equation}\label{Eq:ExpectA-final}
\Braket{A} = \sum_n \Braket{n|A \rho|n} = \text{Tr}(\rho A).
\end{equation}
$\rho$ is the density operator, which contains probabilities $w_i$  of finding the population in state $\Ket{i}$.

We call the set $\{w_i\}$ {\em classical probabilities} because they represent the fraction of systems within an ensemble that are in state $\Ket{i}$, a classical concept.  The coefficients $a_i$ in the state $\Ket{\psi} = \sum_i a_i \Ket{i}$ give what we call the {\em quantum probabilities} $\left| a_i \right|^2$ because they arise from superposition, a quantum mechanical property.

The density operator in this form was first introduced by von Neumann,\cite{neuma27.01} who explains the need for the classical probabilities, ``{\em But such statistical considerations acquire a new aspect when we do not even know what state is actually present...}"\cite{neuma18.01}

The above pedagogical derivation can be found at least as far back as in Richard Tolman's 1938 book, where he suggests the need for ``{\em \ldots taking the mean first over the range of possibilities presented by each member of the ensemble and then over the members of the ensemble,}" where the ``possibilities" refers to quantum probabilities and the ``members of the ensemble" the classical average from the distribution of populations.\cite{tolma38.01}

In 1957, a review article by Ugo Fano reinforces this sentiment by explaining\cite{fano57.01} ``{\em States that are not `pure' have been called `mixed' states because they can be described by the incoherent superposition of pure states. Incoherent superposition means, by definition, that to calculate the probability of finding a certain experimental result with a system in the mixed state one must first calculate the probability for each of the pure states and then take an average, attributing to each of the pure states an assigned `weight.'} "

This development again mixes classical and quantum concepts.  The state vectors are quantum mechanical in nature but the separation into populations is a classical construct.  Landau warns us that the universe is governed by quantum mechanics and viewing this separation classically is wrong.\cite{landa69.01}  In his own words, ``{\em \ldots it would be quite incorrect to suppose that the description by means of the density matrix signifies that the subsystem can be in various $\Psi$ states with various probabilities and that averaging is over these probabilities.  Such a treatment would be in conflict with the basic principles of quantum mechanics.}"  He then continues by emphasizing that, ``{\em \ldots the statistical averaging [is] necessitated by the incompleteness of our information concerning the objects considered.}"  Subsequently, he stresses that the two probabilities cannot be separated and that ``{\em the whole averaging procedure is carried out as a single operation, and cannot be represented as the result of successive averaging, one purely quantum-mechanical and the other purely statistical.}"  Many standard textbooks present the student with an explanation that does not heed Landau's warning.  Band and Park disagree with this all-quantum view, arguing that measurements are often repeated many times, so the statistics resides in the variability of preparing the system.\cite{band79.01}

Landau and Lifshitz (L\&L) present what appears to be a more precise derivation of the density operator along the following lines\cite{landa77.01} by defining $\psi(x,q)$ to be the wavefunction of the universe, which can be formally calculated from the state vector $\ket{\psi}$ of the universe using the inner product $\psi(x,q) = \braket{xq| \psi}$.  $\left| \psi(x,q) \right|^2$ is understood to be the probability density of finding the object in a state defined by the parameter $x$ and a particle in the environment described by parameter $q$.  Since $x$ and $q$ can each describe many degrees of freedom, this wavefunction depends on $x = (x_1, x_2, \dots)$ in the object and $q = (q_1, q_2, \dots)$ in the environment.   The wavefunction is usually not separable, so that $\psi(x,q) \neq \psi(x) \psi(q)$.

The expectation of the operator $A$ is given by
\begin{align}\label{eq:expectA}
\Braket{A} = \int dx \int dq \,  \psi^* (x,q) A(x,q) \psi(x,q) .
\end{align}
Since only the system is being measured, L\&L argue that the operator $A$ is restricted to parameters $x$.  Then, $A(x,q) \rightarrow A(x)$ -- a somewhat confusing assumption, which we clarify later.  With A(x) depending only on $x$, Equation \ref{eq:expectA} after some regrouping of terms becomes
\begin{align}\label{eq:expectAdensity}
\Braket{A}_x = \int dx \, A(x) \int dq \,  \psi^* (x,q) \psi(x,q) \equiv \int dx \, A(x)\rho(x),
\end{align}
where $\rho(x)$ is the diagonal part of the density matrix, which is defined by
\begin{align}\label{eq:DensityMatrixDefine}
\rho(x, x^\prime) = \int dq \,  \psi^* (x,q) \psi(x^\prime,q) .
\end{align}
Clearly, $\rho^*(x, x^\prime) = \rho(x^\prime , x)$ so the density matrix represents a hermitian operator.

The diagonal elements of the density matrix are the probability densities of finding the system with parameter $x$ and are of quantum origin.  It is impossible to know the state of the environment,  which is entangled with the system, so the density matrix is modeled with phenomenological parameters $w_i$ in Equation \ref{Eq:DensityOp}, which have the feel of classical probabilities.  Our simple two-particle model will provide a simple example of how this works.

The L\&L derivation does not provide a clear way of getting the density matrix since $\psi(x,q)$ includes the coordinates of the environment, whose effects are too complex to take into account.  As we show later, defining a system by its spatial region leads to additional confusion for identical particles, which must be entangled.  At the other extreme, one can use Equation \ref{Eq:DensityOp} to interpret the coefficients $w_i$ as populations, which as we have seen is justifiably objectionable to Landau.  Feynman's approach follows Landau's development,\cite{feynm72.01} but connects the quantum calculation to the classical interpretation more clearly.  However, Feynman's book lacks simple examples to explain how the universe is parsed and contains the seemingly erroneous statement that $\Ket{i}$ must generally form a compete orthonormal set of state vectors.

Our goal is to use simple examples that show the student how the density matrix emerges from an entangled state and to clear up confusion that might arise from textbooks and the literature.  In a nutshell, this paper can be summarized by the following example of how a density matrix representing a mixed state emerges from two non-interacting electrons with state vector
\begin{align} \label{eq:two-electron-state}
\Ket{\psi} = \frac {1} {\sqrt{2}} \Big( \Ket{\uparrow} \ket{\downarrow} - \ket{\downarrow} \Ket{\uparrow} \Big).
\end{align}
Landau first noticed the emergence of mixed states in this way.\cite{landau27.01} For an operator $A_1$ that acts only on the first particle (leftmost ket in each term), its expectation value is given by
\begin{align} \label{eq:two-electron-A1-expect}
\Braket {\psi | A_1 | \psi} & = \frac {1} {2} \Big( \Braket{\uparrow|A_1|\uparrow} \Braket{\downarrow|\downarrow} + \Braket{\downarrow|A_1|\downarrow} \Braket{\uparrow|\uparrow} \nonumber \\ & - \Braket{\uparrow|A_1|\downarrow} \Braket{\downarrow|\uparrow} - \Braket{\downarrow|A_1|\uparrow} \Braket{\uparrow|\downarrow}\Big) \nonumber \\
& = \frac {1} {2} \Big( \Braket{\uparrow|A_1|\uparrow} + \Braket{\downarrow|A_1|\downarrow} \Big).
\end{align}
Equation \ref{eq:two-electron-A1-expect} is equivalent to Equation \ref{Eq:ExpectA-final} with $A_1 = A$ and a density matrix given by
\begin{align}\label{eq:two-spin-denisty-matrix}
\rho = \left(
         \begin{array}{cc}
           \frac {1} {2} & 0 \\
           0 & \frac {1} {2} \\
         \end{array}
       \right)
\end{align}
-- a mixed stated.

This simple example shows that:
\begin{enumerate}
\item The progenitor of a one-particle mixed state can be a pure state (in this case, two entangled particles).\label{list:progenitor}
\item The coefficients $w_i$ originate from the coefficients of superposition in Equation \ref{eq:two-electron-state}, which represent quantum probability amplitudes, i.e.~the factors $1/\sqrt{2}$.\label{list:QunatProb}
\item Mixed states do not require classical populations.\label{list:pure-to-mix}
\item A mixed state emerges when measuring an observable whose associated operator acts on a one-particle subset of a pure and entangled two-particle state.\label{list:particleSub}
\end{enumerate}

These observations can be generalized to many particles, but the important points can be made with two particles alone.  Lerner showed in a similar way how decoherence can be understood without a density matrix\cite{lerne17.01} and Press avoided the density matrix to glean insights that are lost about entangled spins when using the density matrix.  Asher, on the other hand, uses arguments similar to ours using two spins but calls upon the time evolution of the density matrix.\cite{peres74.01} von Neumann's excellent book, which has been typeset anew and published in 2018 provides many insights on projections into subsystems,\cite{neuma18.01} but at a level more advanced than the intended audience of this paper.

This view is not new and the quantum information theory literature commonly accepts the fact that the mixed state is part of a larger purely quantum system.  In fact, it is a simple matter to find a many-particle pure state that gives a particular density matrix through a process called ``purification,"  and there are practical ways to implement purification.\cite{klein06.01,Nielsen01.10}  However, the more specialized textbooks do not make such connections clear at an elementary level, and are not used in core physics classes.  For example, Nielsen and Chuang start with ``{\em suppose that a quantum system is in one of a number of states $\Ket{ \psi_i}$, where $i$ is an index, with respective probabilities $p_i$}" then they define the density matrix by $\sum_i p_i \Ket{\psi_i} \Bra{\psi_i}$ with no explanation, taking for granted that the reader already has an understanding of the topic.\cite{Nielsen01.10}  A significant effort would be required of a beginning quantum mechanics student to use such texts to come to the understanding embodied in the simple example leading to Eq. \ref{eq:two-spin-denisty-matrix}.

One motivation for this paper is to point out that this connection between pure quantum states and density matrices should be made when the topic is first introduced to students, and to urge instructors to consider doing so using simple two-particle models.  A central theme of this paper is to emphasize that the richness of phenomena described by density operators, such as mixed states and the time evolution of populations, can be understood using an entangled pair of two particles.

In the sections that follow, we start by defining what is meant by an object and environment, then consider such universes made of distinct then identical particles.  Interactions between particles are shown to be required to get density matrices with diagonal components that decay exponentially over time to their equilibrium values, as is observed in experiments and phenomenologically modeled with the time evolution of density matrices.  Most importantly, we show how the coefficients used in quantum superposition are related to the density matrix elements, thus providing insights into the connection between the two.  We use these observations to highlight common misconceptions and the confusion that might result.

\section{Universe as an Object in an Environment}\label{sec:SysUni}

\begin{figure}\centering
    \includegraphics{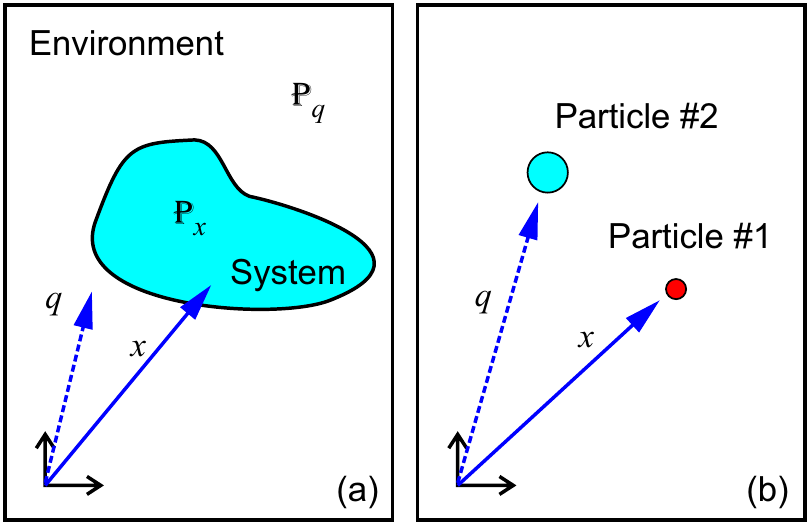}
    \caption{(a) A position coordinate within the universe is labelled $x$ when inside the system under study and $q$ when in the environment.  $x$ and $q$ can represent many degrees of freedom. (b) Alternatively, the system can be defined to be Particle \#1 (or a collection of particles) with coordinate(s) $x$ and the environment as Particle \#2 (or a collection of particles) with coordinate(s) $q$.  In this case $x$ and $q$ can occupy the same spatial region.}
    \label{fig:System-etc}
\end{figure}
There are two common ways to divide the universe into a system/object and environment.  Figure~\ref{fig:System-etc}a shows a spatially-isolated system.  Then, the environment is the part of the universe that falls outside the system.  In this case, $x$ and $q$ label the coordinates inside the object and in the environment, respectively.  Alternatively, as shown in Figure~\ref{fig:System-etc}b, the system is not defined by the region it occupies, but by the type of particle -- for example a large blue one or a small red one.  This makes it possible to ask ``What is the probability of finding the red particle at position $x$?"  In contrast, for indistinguishable particles, one can only ask ``What is the probability of finding a particle at position $x$?" without specifying which one.

The following subsection starts with the simplest example of two distinguishable particles to illustrate how the density matrix is constructed for one of the particles as the system, and the other particle as the environment.  The concluding subsection derives the density matrix when the system and environment are two spatially separated regions.  The two regions can be viewed as two wells that each contain a particle, thus forming a two-particle pure state. {\em In the presence of infinite wells}, either method of identification is equivalent and will yield the same result, though splitting the universe into two particles by type is more intuitive than spatially separating them when including spin statistics.\cite{kuzyk19.01}

\subsection{Particle as System}

The most general two-particle wave function is of the form
\begin{align}\label{eq:2-particle-General-energy-rep}
\Ket{\psi} =  \sum_{n_1 , n_2} c_{n_1 , n_2}\Ket{n_1} \Ket{n_2} ,
\end{align}
where the subscripts $1$ and $2$ are the particle labels.  Then, the expectation value of the observable $A_1$ of Particle \#1 is given by
\begin{align}\label{eq:2-particle-energy-rep-A1}
\Braket{\psi|A_1|\psi} =  \sum_{n_1 , n_2} \sum_{n_1^\prime , n_2^\prime} c_{n_1 , n_2}^*  c_{n_1^\prime , n_2^\prime} \Braket{n_1 | A_1 | n_1^\prime} \delta_{n_2 , n_2^\prime} .
\end{align}

Defining
\begin{align}\label{eq:2-particle-energy-rep-density}
\rho_{n_1 , n_1^\prime} =  \sum_{n_2} c_{n_1 , n_2}^*  c_{n_1^\prime , n_2} ,
\end{align}
Equation \ref{eq:2-particle-energy-rep-A1} becomes
\begin{align}\label{eq:2-particle-energy-rep-A1-den}
\Braket{\psi|A_1|\psi} =  \sum_{n_1 , n_1^\prime} \rho_{n_1 , n_1^\prime} \Braket{n_1 | A_1 | n_1^\prime} = \text{Tr}(\rho A_1).
\end{align}
Equation \ref{eq:2-particle-energy-rep-A1-den} is equivalent to Equation \ref{Eq:ExpectA-final},  so the expectation value of the ``system" originates from a two-particle pure state in a {\em quantum} superposition, but can be expressed in terms of a density matrix, which depends on the coefficients $c_{n_1,n_2}$.

The operator form of Equation \ref{eq:2-particle-energy-rep-density} is
\begin{align}\label{eq:2-particle-energy-rep-density-op}
\rho =  \sum_{n_2} \Braket{n_2 | \psi} \Braket{\psi | n_2} ,
\end{align}
where the density operator given by Equation \ref{eq:2-particle-energy-rep-density-op} of the object is the part that is left over after taking the trace over the environment of the pure state operator of the universe $\Ket{\psi} \Bra{\psi}$.  Convince yourselves that this is true by calculating $\Braket{n_1^\prime | \rho | n_1}$ of Equation \ref{eq:2-particle-energy-rep-density-op} with the help of Equation \ref{eq:2-particle-General-energy-rep} to get Equation \ref{eq:2-particle-energy-rep-density}.

I leave it as an exercise for the student to use Equation \ref{eq:2-particle-energy-rep-density} to show that the density matrix is hermitian and has unit trace.  Hermiticity and unit trace allows the density matrix to be diagonalized to yield an expression in the form of Equation \ref{Eq:DensityOp} with what appear to be classical population fractions  $w_i$ that are related to the $c_{n_1 , n_2}$ coefficients of Equation \ref{eq:2-particle-energy-rep-density}.  While this diagonal form has the appearance of classical probabilities given by $w_i$, their origin here is purely quantum in nature.

Viewing a system as being made of an ensemble, whose properties are determined with statistical weighting, has been a highly successful paradigm for building models of a wide variety of systems and phenomena that result.  However, one should not lose sight of the fact that our universe need not naturally divide into little quantum compartments that form a larger collection of classical populations.

\subsection{Spatial boundary between Object and Environment}

To treat spatial regions, we label the coordinates of the distinguishable particles $r_1$ and $r_2$, where $r_i$ is given by $x_i$ inside the object and $q_i$ outside.  We restrict the calculation that follows to the case where each particle is fully localized in its own well.

For a universe described fully by the state vector $\Ket{\psi}$, the probability amplitude of finding Particle \#1 in the system at position $x_1$ and Particle \#2 in the environment at position $q_2$ is given by
\begin{align}\label{eq:2-particle-prob}
\psi(x_1, q_2) =  \Bra{x_1} \Bra{q_2} \cdot \Ket{\psi}.
\end{align}
Therefore, in the position representation, the two-particle state vector is
\begin{align}\label{eq:2-particle-General-position-rep}
\Ket{\psi} =  \int dq_2 \, \int dx_1 \, \psi (x_1, q_2) \Ket{x_1} \Ket{q_2}  .
\end{align}
The student should apply $\Bra{x_1^\prime} \Bra{q_2^\prime}$ to both sides of Equation \ref{eq:2-particle-General-position-rep} to verify that Equation \ref{eq:2-particle-prob} results. Equation \ref{eq:2-particle-General-position-rep} is the continuous form of Equation \ref{eq:2-particle-General-energy-rep}.  The expectation of observable $A$ of the system is then
\begin{align}\label{eq:2-particle-General-position-A}
& \Braket{\psi|A|\psi} = \int dq_2^\prime \, \int dx_1^\prime \, \int dq_2 \, \int dx_1 \, \psi^* (x_1^\prime, q_2^\prime) \nonumber \\
& \times \psi (x_1,q_2)  \Braket{x_1^\prime|A|x_1} \braket{q_2^\prime|q_2} \\ \nonumber
&= \int dx_1^\prime \, \int dq_2 \, \int dx_1 \, \psi^* (x_1^\prime, q_2) \psi (x_1,q_2) \Braket{x_1^\prime|A|x_1}
,
\end{align}
which is a more general result than that of Landau and Lifshitz given by Equation \ref{eq:expectAdensity} but reduces to their result when $\Braket{x_1^\prime | A | x_1} = \delta (x_1 - x_1^\prime) A(x_1)$.  {\em We stress that the derivation leading to Equation \ref{eq:2-particle-General-position-A} assumes that the single-particle wavefunctions are each confined to one well or the other one, a way in which the universe is typically separated, but an idealization of reality.\cite{kuzyk19.01}}

Equation \ref{eq:2-particle-General-position-A} can be written as
\begin{align}\label{eq:2-particle-General-position-A-den}
\Braket{\psi|A|\psi} = \int dx^\prime \int dx \, \rho(x, x^\prime) \Braket{x^\prime|A|x},
\end{align}
where $\rho(x^\prime,x)$ is given by Equation \ref{eq:DensityMatrixDefine}.  Here we have dropped the subscript to label the particle because we know that Particle \#1 is in Well \#1 with spatial coordinates $x$.

Equation \ref{eq:2-particle-General-position-A} makes clear that the double integral over the environment collapses into a single integral due to the inner product $\Braket{q_2^\prime|q_2}$.  In contrast, Landau states Eq. \ref{eq:expectA} without proof, which is an obvious fact to experts but not to students learning the material for the first time.  Furthermore, Equation \ref{eq:2-particle-General-position-A} is a more general result that allows for off-diagonal elements of the operator $A$.

To summarize, a universe with two distinguishable particles in two separate infinite wells can be referenced in two ways; if Particle \#$i$ is placed in Well \#$i$, then the particle label is the same as the well label.  Then, the projector onto Particle \#i is the same as the projector into Well \#i.  Then
\begin{align}\label{eq:SameProjector1}
\sum_{n_1} \Ket{n_1} \Bra{n_1} = \int_{\mbox{Well} \#1} dx \, \Ket{x} \Bra{x}
\end{align}
and
\begin{align}\label{eq:SameProjector2}
\sum_{n_2} \Ket{n_2} \Bra{n_2} = \int_{\mbox{Well} \#2} dq \, \Ket{q} \Bra{q} .
\end{align}

In this example, each well is in a different physical subspace -- one here and one there.  Particles, on the other hand, always live in different Hilbert spaces even when they occupy the same physical space.  In the special case where each particle is confined to its own infinite well, thus defining two distinct regions of physical space, there is a direct correspondence between the Hilbert space and the physical space, so they can be used interchangeably.  This correspondence does not hold in the absence of such localized wavefunctions, where the Hilbert space is unrelated to the physical space.  Pedagogically, it is preferable to parse the universe by particle type, because it clearly shows that Hilbert space is not the same as physical space.\cite{kuzyk19.01}

To summarize, this section has shown that when the system and the environment is defined by two regions of space or by two distinguishable particles, the expectation values of an operator for one of the particles takes the form given by Equation \ref{Eq:ExpectA-final}, where the density matrix is a function of the coefficients from the superposition of states that defines the universe.

\section{Interacting Indistinguishable Particles}

This section shows how the density matrix of one particle or one confined region of space emerges from the pure state of {\em two indistinguishable fermions or bosons}, which must obey the spin statistics theorem so therefore must be entangled.\cite{duck98.01}  Spatial entanglement is an interesting topic, but applying it here adds no new physics to the arguments that follow.\cite{schro17.01}

\subsection{Two-Particle State Vector Basis}

If two particles are indistinguishable, we cannot label them individually so the wave functions must be either symmetric or antisymmetric upon exchanging them.  Thus, we can only compute the probability amplitude of finding one particle in one state and one particle in another one without referencing which one is which.  In Dirac notation, the state vector of two noninteracting particles that {\em occupy any two of the single-particle eigenstates} $\Ket{n}$ and $\Ket{m}$ is of the form
\begin{align}\label{eq:2-particle-energy_eigen}
\Ket{n,m} = \frac {1} {\sqrt{2}} \Big[ \Ket{n} \Ket{m} \pm \Ket{m} \Ket{n} \Big].
\end{align}

Note that two kets together such as $\Ket{n} \Ket{m}$ represent a two-particle state, with Particle \#1 in a one-particle eigenstate $\Ket{n}$ with eigenvalue $n$ and Particle \#2 in a one-particle eigenstate $\ket{m}$ with eigenvalue $m$.  The eignevalue associated with $n$ could, for example, be the energy $E_n$, but consider it to be general.

The first ket in a pair refers to Particle \#1 and the second ket to Particle \#2.  Each of the two terms in Equation \ref{eq:2-particle-energy_eigen} distinguish between the two particles, yet when added together, the state $\Ket{n,m}$ does not;  that is, state $\Ket{n,m}$ corresponds to each particle being in both states.  This result applies to multiple particles in one well or particles in two different wells.

Equation \ref{eq:2-particle-energy_eigen} in position space is given by
\begin{align}\label{eq:2-particle-energy-wavefucntion}
\psi(x,q) &= \Braket{x,q| n,m} = \frac {1} {2} \Big[ \psi_n(x) \psi_m (q) \pm \psi_n (q) \psi_m(x)  \nonumber \\
&+ \psi_m(x) \psi_n (q) \pm \psi_m (q) \psi_n (x) \Big],
\end{align}
where $\Ket{x,q}$ is in the same (anti)symmetrized form as Equation \ref{eq:2-particle-energy_eigen}, so each particle lives in both the object and the environment.  {\em Note that we have kept the wavefunctions ordered to keep track of which particle is in which state and in which region.}  For example, $\psi_m(x) \psi_n (q)$ means that Particle \#1 is in state $m$ and inside the object (position $x$) while particle \#2 is in state $n$ and in the environment (position $q$).

Let's pick two wells that have an irrational ratio of widths so that the single-particle energy eigenfunction vanishes in one well when it is nonzero in the other one.  Since one particle occupies each well, let's arbitrarily pick state $n$ to be found in the object and state $m$ in the environment.  Then, $\psi_n(q) = \psi_m(x) = 0$ because state $n$ vanishes in the environment and state $m$ vanishes in the object.  Equation \ref{eq:2-particle-energy-wavefucntion} then becomes
\begin{align}\label{eq:2-particle-energy-wavefucntion-vanish}
\psi(x,q) &=  \frac {1} {\sqrt{2}} \Big[ \psi_n^{(1)}(x) \psi_m^{(2)} (q) \pm \psi_m^{(1)} (q) \psi_n^{(2)} (x) \Big],
\end{align}
where we have labeled each particle's identity as a superscript to remind us which wavefunction is associated with which particle.

We note that the wavefunction given by equation \ref{eq:2-particle-energy-wavefucntion-vanish} associates an energy eigenstate index ($n$ or $m$) with a particular well ($x$ or $q$).  In contrast to the previous case where particle type is associated with a spatial region, here the particle types are indistinguishable.  We arbitrarily associate the energy index $n$ with the spatial region defined by $x$ and $m$ with $q$.  Thus, all of the previous discussions for distinguishable particles caries forward here by replacing ``particle type" with ``energy index."  We need not rehash these discussions for indistinguishable particles.  Instead, we will focus on the particle as the object for a system of two indistinguishable particles, but will add interactions between them to show that this will lead to all possible density matrices.

\subsubsection{A Particle as the Object}

When two particles interact with each other, the energy eigenstate vectors $\Ket{\psi^{(k)}}$ of the full Hamiltonian
\begin{equation}\label{Eq:TwoParticleH}
H = H_1 + H_2 + V({\bf r}_1,{\bf r}_2)
\end{equation}
can be expressed as the superposition
\begin{equation}\label{Eq:TwoParticleEnergyEigen}
\Ket{\psi^{(k)}} = \frac {1} {\sqrt{2}}\sum_{n,m} c_{n,m}^{(k)} \Ket{nm} ,
\end{equation}
where we have used the fact that the set of state vectors $\left\{ \Ket{nm} \right\}$ form an orthonormal basis. For bosons/fermions, since $\ket{nm} = \pm \ket{mn}$, we choose the complex coefficients to obey $c_{n,m}^{(k)} = \pm c_{m,n}^{(k)}$.  For each state $k$, normalization requires
\begin{equation}\label{eq:Norm}
\sum_{n,m} \left| c_{n,m}^{(k)} \right|^2 = 1 .
\end{equation}
Alternatively, we are free to restrict the sums in Equation \ref{Eq:TwoParticleEnergyEigen} to terms with $m > n$ for fermions (since $\Ket{nm} = 0$) and $m \ge n$ for bosons.   We can choose either form based on convenience.

The $k^\text{th}$ eigen energy $E^{(k)}$ is given by
\begin{equation}\label{Eq:TwoParticleEnergy}
H \Ket{\psi^{(k)}} = E^{(k)} \Ket{\psi^{(k)}} .
\end{equation}
Note that $H_i$ in Equation \ref{Eq:TwoParticleH} is the one-particle Hamiltonian of Particle \#i so $H_1 \Ket{n} \Ket{m} = E_n \Ket{n} \Ket{m} $ and yields $\left(H_1 + H_2 \right) \Ket{nm} = \left(E_n + E_m \right) \Ket{nm} $.  The student should verify this relationship.

The most general state vector, then, will be a superposition of the two-particle state vectors given by Equation \ref{Eq:TwoParticleEnergyEigen}, yielding
\begin{equation}\label{Eq:TwoParticleSuperpose}
\Ket{\psi} = \sum_k a_k \Ket{\psi^{(k)}} = \frac {1} {\sqrt{2}} \sum_k a_k \sum_{n,m} c_{n,m}^{(k)} \Ket{nm} ,
\end{equation}
where $\sum_k \left| a_k\right|^2 = 1$ to maintain normalization.  The expectation value of operator $A_1$, which acts only on particle \#1, yields
\begin{align}\label{Eq:TwoParticleSuperposeA1expect}
\Braket{\psi|A_1| \psi} & = \frac {1} {2} \sum_{k,k^\prime} a_{k^\prime}^*a_k \sum_{n^\prime,m^\prime} \sum_{n,m}  c_{n^\prime,m^\prime}^{(k^\prime)*} c_{n,m}^{(k)} \nonumber \\
& \times \Braket{n^\prime, m^\prime| A_1 | n, m}
\end{align}

Re-expressing Eq. \ref{Eq:TwoParticleSuperposeA1expect} in terms of the single-particle states using Eq. \ref{eq:2-particle-energy_eigen}, and recognizing that only the first bras and kets in the pair act on $A_1$ and the second ones pass through -- as illustrated in Eq. \ref{eq:two-electron-A1-expect}, we get
\begin{align}\label{eq:2-particle-General-Superpose-Expect-single}
\Braket{\psi|A_1|\psi} & = \sum_{k,k^\prime} a_{k^\prime}^*a_k  \sum_{n^\prime,m^\prime} \sum_{n,m} \frac { c_{n^\prime,m^\prime}^{(k^\prime)*} c_{n,m}^{(k)}} {4} \nonumber \\
& \Big( \Braket{n^\prime | A_1 | n}  \delta_{m,m^\prime} + \Braket{m^\prime | A_1 | m}  \delta_{n,n^\prime} \nonumber \\
& \pm \Braket{n^\prime | A_1 | m}  \delta_{n,m^\prime} \pm \Braket{m^\prime | A_1 | n}  \delta_{m,n^\prime} \Big) ,
\end{align}
where the $\pm$ signs are for bosons and fermions.

Next we combine terms in Equation \ref{eq:2-particle-General-Superpose-Expect-single} by changing the dummy indices in the sums and using the fact that $c_{nm}^{(k)} = \pm c_{mn}^{(k)}$, which is a consequence of $\Ket{n,m} = \pm \Ket{m,n}$.  For example, the second term in parentheses in Equation \ref{eq:2-particle-General-Superpose-Expect-single} with its pre-factor $c_{n^\prime,m^\prime}^{(k^\prime)*} c_{n,m}^{(k)}$ can be converted into the first term by  interchanging the dummy index $m^\prime \rightleftharpoons n^\prime$ then interchanging the indices $m \rightleftharpoons n$.  It is left as an exercise for the student to manipulate the other terms in this way, which yields
\begin{align}\label{eq:2-particle-General-Superpose-Expect-single-2}
\Braket{\psi|A_1|\psi} & = \sum_{k,k^\prime} a_{k^\prime}^*a_k \sum_{n^\prime} \sum_{n,m}  c_{n^\prime,m^\prime}^{(k^\prime)*} c_{n,m}^{(k)}  \Braket{n^\prime | A_1 | n} \nonumber \\
& \equiv \sum_{n,n^\prime} \rho_{n, n^\prime} \Braket{n^\prime | A_1 | n} = \mbox{TR[$\rho A_1$]},
\end{align}
where the density matrix is given by
\begin{align}\label{eq:2-particle-General-Interact-Density}
\rho_{n^\prime , n} = \sum_{k^\prime , k} a_{k^\prime}^* a_k \sum_m c_{n^\prime,m}^{(k^\prime) *} c_{n,m}^{(k)} .
\end{align}
Equation \ref{eq:2-particle-General-Interact-Density} is the most general form resulting from two interacting particles that obey the spin statistics theorem.\cite{duck98.01}  This two-interacting-particle state vector yields the general form of the density matrix, where its matrix elements originate in the coefficients of the superposition expansion given by Equation \ref{Eq:TwoParticleSuperpose}.

As an example of imprecise language that implies the density matrix' ability to account for phenomena beyond the quantum realm, consider Blum's statement on page 7 in his book {\em Density Matrix Theory and Applications} when referring to a beam of spin-1/2 particles,\cite{blum12.01} ``{\em Clearly it is not possible to characterize the state of the beam in terms of a single state vector $\Ket{\chi}$ since associated with any of these states there is necessarily a direction in which {\em all} spins are pointing; the direction of the polarization vector.  If the Stern-Gerlach filter were placed in this orientation the whole beam would have to be transmitted.  Since no such orientation exists it is not possible to describe a mixture by a single state vector.}"  Articles in this journal repeat such arguments with even stronger statements such as ``{\em ...hence it cannot be represented by {\em any state vector},}"\cite{peres74.01} my emphasis.

We derived Equation \ref{eq:2-particle-General-Interact-Density} from a single state vector made of just two entangled particles, and it is the same as Blum's Equation 2.11.  Hence, we have found a pure state of two entangled particles in superposition with properties that Blum and others imply are impossible for a pure state.  Thus, the density matrix {\bf is not} more general than a two-particle pure state.  Statements of the sort quoted in the previous paragraph are not true and confusing to a student who notices the inconsistency while trying to learn about density matrices.  Most often, such contradictions reinforce the notion that density matrices are in some way more general than state vectors and mask the simple fact that all these properties emerge form a pure state that is represented by a state vector.

Blum's narrative is commonly found in textbooks, lectures and the literature -- which leaves the impression that only a density matrix will do.  In reality, an experimental observation is {\em most conveniently characterized with a density matrix} because an infinite number of progenitor pure states lead to the same observation, making it impossible to determine the unique one for that particular system.  Even within a given basis, there are an infinite number of ways to construct a particular density matrix by virtue of the fact that mixtures can include non-orthogonal states.  However, using Equation \ref{eq:2-particle-General-Interact-Density}, a two-particle state vector can be identified that in the one-particle subspace fully describes any experimental observation of that system, so can be used in lieu of the density matrix.

The density operator for indistinguishable particles cannot be determined by inspection as it was in obtaining Equation \ref{eq:2-particle-energy-rep-density-op} for distinguishable particles because indistinguishable-particle wave functions are entangled.  However, the matrix elements $\rho_{n,n^\prime}$ given by Equation \ref{eq:2-particle-General-Interact-Density} can be used to construct the density operator using
\begin{align}\label{eq:2-particle-energy-rep-density-op-indist}
\rho =  \sum_{n, n^\prime} \Ket{n} \rho_{n, n^\prime} \Bra{n^\prime},
\end{align}
which can be diagonalized to get Equation \ref{Eq:DensityOp} by virtue of its hermiticity and unit trace.

\section{Discussion}\label{sec:Discussion}

This section applies the general results presented in prior sections to two important examples that are often used to argue for the need of density matrices representing mixed states in lieu of state vectors.  These include (1) the Stern Gerlach experiment, which in certain preparations finds the signal to be independent of magnetic field direction and (2) the fact that many processes -- such as population decay -- are irreversible and thus not describable by unitary time evolution.  This section illustrates that a simple two-particle pure state can have these properties.  Our aim is to show that such behavior can result, and not to provide a many-particle calculation for a real system, countering the suggestion of many texts that a density matrix is required.

\subsection{Spin Mixtures}\label{sec:SpinMix}

Fano argues that density matrices are needed when information about a system is lacking.  He uses as an example a particular preparation of an atom beam measured by a Stern-Gerlach apparatus that yields a vanishing expectation value of the spin along any arbitrary axis.  He then asserts that a pure state cannot explain this result. Specifically, he states that , ``{\em \ldots lack of information about a variable can often be expressed as a statement that certain operators have expectation value zero. For example if the spin orientation of a particle is wholly unknown, the expectation value of each component of its angular momentum vanishes.}"  We offer Equation \ref{eq:two-electron-state} as a counterexample. It is a well-defined pure state that yields vanishing expectation values of the spin of each of the two particles.

Using the state vector defined in Equation \ref{eq:two-electron-state} the student can use Equation \ref{eq:two-electron-A1-expect} with $A_1$ being the one-particle spin operator to verify that the expectation value of the spin of the $n^\text{th}$ electron is
\begin{align}\label{eq:spin-epxect}
{\bf S}^{(n)} = \Braket{\psi| \sum_{p}   S_p^{(n)} \hat{p} | \psi} = 0,
\end{align}
where $S_p^{(n)}$ is the $p^\text{th}$ cartesian component of the spin vector of the $n^\text{th}$ electron.  The same result will be obtained independent of the coordinate system.  ${\bf S}^{(n)} = 0$ satisfies Fano's requirement, which he claims implies that the state of the system is unknown, yet we have shown that his requirement is met with a well-defined two-particle state vector.

This example presents a teachable moment for how the density matrix can be presented without giving the impression that the density matrix is an extension of quantum mechanics.  One possible set of exercises follows.
\begin{enumerate}

\item Have the student calculate the expectation value of the spin, let's say $S_z$, for the general single-particle state $\psi(\alpha) = \cos \alpha \Ket{\uparrow} + e^{i \phi} \sin \alpha \Ket{\downarrow} $ as a function of $\alpha$ to illustrate that the expectation value depends on $\alpha$.  Alternatively, the state vector can be fixed and the spin operator rotated to show that the expectation values agree with the behavior of projections of classical spins.

\item Have the student calculate the expectation value of the spin of one particle in an entangled pair using the state vector given by Equation \ref{eq:two-electron-state} to show that the spin vanishes for all orientations of the single-particle spin operator.  This illustrates that the expectation value of one particle that is part of a collection of many particles behaves quite differently than a particle in isolation. \label{Item:Entangle}

\item Starting from the density matrix given by Equation \ref{eq:two-spin-denisty-matrix}, show that the expectation value of the one-particle spin is the same as that given by Item \ref{Item:Entangle}.

\item Explain to the student that the density matrix can be interpreted as describing a population of spins, where half of them are spin up and half of them are spin down.  Stress that this property emerges from a pure state vector of entangled particles, so the system is fully describable by quantum mechanics, but that it is often convenient to use a density matrix rather than a state vector for a many particle system.
\end{enumerate}

After dong the above exercises, the student will appreciate that an experimentalist measuring an expectation value of ${\bf S} = 0$ would conclude that the density matrix is given by Equation \ref{eq:two-spin-denisty-matrix} but would have no knowledge of the progenitor pure state.  A more accurate statement than Fano's is that {\em a particular density matrix determined from an experiment can correspond to many possible pure states, but that does not imply that no such pure state exists}; a pure state is unique but a density matrix can be the offspring of an infinite number of pure states.

This example dispels the notion that the density matrix embeds more information than the state vector and that quantum mechanics is in some ways lacking.  Weinberg suggests that the density matrix might be the fundamental object of nature and that physics can be described without state vectors.\cite{weinb14.01}  Since any density matrix of a one-particle subsystem can be constructed from a superposition of two-particle states, the state vector and density matrix for such a system are mathematically equivalent as long as we keep in mind that the system is part of a larger universe.  Thus, one is no more fundamental than the other.  This view is often referred to as ``{\em Church of the Larger Hilbert Space,}" following John Smolin of IBM.\cite{gottes00.01}

\subsection{Time Evolution}\label{sec:TimeEvolution}

This section shows that a pure two-particle state in an entangled stated and in superposition can yield what appears to be non-unitary time evolution.  In particular, we answer the question of how an exponential population decay is compatible with the unitary time evolution of a pure state, where probability amplitudes normally oscillate in time.  The research literature is filled with examples where ``Breakdown in unitarity can lead to exponential behavior..."\cite{Beau17.01}  Our narrative shows that a breakdown of unitarity is not required.

Physicists model exponential population decay by adding a phenomenological parameter to the time evolution equations of the density matrix. At first glance, it would appear that quantum mechanics must be augmented or generalized beyond the traditional theory to get the observed behavior.

In the density matrix formalism, exponential population decay originates in the diagonal components of the density matrix.  In contrast, the diagonal components of a pure state are time independent.  When the particles interact, the diagonal elements become time dependent.  In a nutshell, we use the one particle subsystem of two interacting particles to find the specific superposition that gives exponential decay.  Why a system would be in that particular superposition is not a question we address here; rather, our aim is to show that an exponential decay is not at odds with the quantum-mechanical description of nature.

The energy eigenvalue equation for interacting particles is given by Equation \ref{Eq:TwoParticleEnergy}.  Applying the time-evolution operator to Equation \ref{eq:2-particle-General-Interact-Density}, which we emphasize comes from a pure two-particle state, yields the time-dependent density matrix
\begin{align}\label{eq:2-particle-General-Interact-Density(t)}
\rho_{n^\prime , n} (t) & = \sum_{k^\prime , k} a_{k^\prime}^* a_k \exp(-i (E_k - E_{k^\prime})t/ \hbar ) \nonumber \\
& \times \sum_m c_{n^\prime,m}^{(k^\prime) *} c_{n,m}^{(k)}  .
\end{align}
The diagonal component $\rho_{n,n} (t)$ is here time dependent and given by
\begin{align}\label{eq:2-exponential-possible?}
\rho_{n, n} (t) & = \sum_{k^\prime , k} A_{k^\prime, k}^{(n)} \exp(i (\omega_{k^\prime} - \omega_k)t ) \stackrel{?}{=} b_n \exp(-\alpha t) ,
\end{align}
where $\omega_k = E_k/\hbar$ and $A_{k^\prime, k}^{(n)} $ is obtained by comparison of Equations \ref{eq:2-exponential-possible?} and \ref{eq:2-particle-General-Interact-Density(t)}, yielding
\begin{align}\label{eq:AkkPrime}
A_{k^\prime, k}^{(n)} = a_{k^\prime}^* a_k \sum_m c_{n,m}^{(k^\prime) *} c_{n,m}^{(k)} ,
\end{align}
with
\begin{align}\label{eq:AkkPrime-Hermetion}
A_{k^\prime, k}^{(n)^*} = A_{k, k^\prime}^{(n)} .
\end{align}

The question then is if the coefficients  $A_{k, k^\prime}^{(n)}$ can be chosen such that the time dependence is an exponential function.  The answer is ``yes."  Appendix \ref{sec:GettingA} shows the derivation for the special case where the energy levels are equally spaced, as one finds in a harmonic oscillator or a particle in free space.  Equation \ref{eq:2-exponential-possible-continuum-integrate} explicitly gives the coefficients.  Thus, one of two interacting particles that are in a superposition of one-particle states can have diagonal density matrix elements that represent exponential population decay.  Asher makes a similar point, arguing that the apparent irreversibility of a measurement resides in the fact that only part of the full state is being observed, but calls upon populations using density matrices.\cite{peres74.01}  Such an exponential decay is usually modelled only with density matrices, but here we see that the full quantum picture suffices, as should be the case for measurements that obey unitary time evolution.\cite{maxwe72.01}  This example closes the loop, showing the student that properties that are beleived to require a density matrix naturally emerge form a pure state.

\subsection{Many Particles}

The most general many-particle state vector can be expanded as a superposition of direct products of single-particle state vectors of the form
\begin{equation}\label{Eq:GenSuperposition}
\Ket{\psi^{(k)}} = \sum_{n,m,\dots,\ell} c_{n,m,\dots,\ell}^{(k)} \Ket{n,m,\dots,\ell},
\end{equation}
where normalization demands that
\begin{equation}\label{Eq:Norm}
\sum_{n,m,\dots,\ell} c_{n,m,\dots,\ell}^{(k)^*} c_{n,m,\dots,\ell} = 1.
\end{equation}
It is not difficult to see that the generalization of Equation \ref{eq:2-particle-General-Interact-Density} to $N$ particles yields
\begin{align}\label{eq:N-particle-General-Interact-Density}
\rho_{n^\prime , n} = \sum_{k^\prime , k} a_{k^\prime}^* a_k \sum_{m, \dots , \ell} c_{n^\prime,m, \dots , \ell}^{(k^\prime) *} c_{n,m, \dots , \ell}^{(k)}  .
\end{align}

There are many ways that a two-particle universe and a many-particle one can give the same density matrix.  We stress again that full knowledge of the density matrix does not allow one to determine a unique progenitor pure state.

\subsection{Multiple-Particle Operators}

All calculations in this paper are limited to the class of one-particle operators.  If we were interested instead in a quantity such as the interaction energy between a pair of particles, we would need to generalize the results to two-particle operators $V_{12}$, which act on the first two kets.  Then, we would need to add at least one more particle to act as the environment.  This type of generalization is messy but straightforward and adds little insight into density matrices.

\section{Conclusion}

We have shown how the density matrix emerges from a purely quantum treatment of a system and environment made of two particles.  What appear to be classical population fractions originate in the superposition coefficients of the purely quantum state vector, illustrating that the density matrix does not imply that the universe is made of a classical collection of populations of tiny quantum systems but that the whole system can be quantum in nature.  Rather, the density matrix is a useful tool for studying systems that are part of a larger universe whose state vector is unknown, but this does not necessarily imply that new physics is required -- as may be interpreted by some students in the way this material is often presented.

We showed how the system and environment can be parsed by particle or region of space, and find that particle parsing leads to less confusion than the more intuitive separation by regions of space.  The two-particle state vector model of the universe as a system of particle interacting with an environment of one particle yields every possible density matrix, showing that density matrices can be a direct consequence of quantum mechanics without need for augmentation.

Finally, the time evolution of the populations is shown to be a consequence of particle interactions.  While this calculation does not prove that the dynamics of a particular many-body quantum system will obey exponential decay to equilibrium, it illustrates how it is possible to get this behavior even in the two-particle case.

Feynman once mused ``{\em It is unknown whether or not the unverse is in a pure state.}"\cite{feynm72.01} We may never know the state of the full universe; but, this paper brings attention to the fact that if mixed states are not required to explain physical phenomena, it is within the realm of possibilities that our universe is in a pure state.  Popescu and coworkers take this view a step further, showing that thermodynamics may be a consequence of measuring a small entangled subspace within a universe that is in a pure state.\cite{Popescu06.01}  The simple two-particle picture, which can be easily generalized to a larger system, gives the student insights into such musings and could serve as an inspiration to learn more.

\vspace{1em}

\noindent{\bf Acknowledgements} We thank the National Science Foundation (EFRI-ODISSEI:1332271 and ECCS-1128076) for generously supporting work that meandered into this unexpected direction.  I am indebted to Mark C. Kuzyk for illuminating discussions, Fred Gittes for insightful suggestions, Ethan Crowell for helpful comments and suggestions, and Daniel Steck for useful references.

\appendix

\section{Summation Gymnastics}

Here we describe useful manipulations of sums.

\subsection{Sums of Differences}\label{sec:sum-of-diff}

Consider the sum
\begin{align}\label{eq:dif-sum}
\sum_{i=0}^{N} \sum_{j=0}^{N} f(i-j),
\end{align}
which is a function of the differences of the indices.  Making the variable change
\begin{align}\label{eq:DeltaDefine}
\delta = i-j
\end{align}
Sum \ref{eq:dif-sum} can be re-expressed as
\begin{align}\label{eq:dif-sum-single}
\sum_{i=0}^{N} \sum_{j=0}^{N} f(i-j) = \sum_{\delta= -N}^N g(\delta) f(\delta) ,
\end{align}
where $g(\delta)$ corresponds to the number of times $\delta = i-j$ appears in the sum.  For example, $\delta = 0$ appears $N+1$ times.  The terms with $\delta = 0$ are $i-j = 0-0, 1-1, \dots , N-N$.  Convince yourselves  that
\begin{align}\label{eq:delta-number-times}
g(\delta) = N + 1- | \delta |
\end{align}
by listing some examples.

Next let's add a coefficient $a_{i,j}$ to the sum, making it of the form
\begin{align}\label{eq:dif-sum-coef}
\sum_{i=0}^{N}\sum_{j=0}^{N} a_{i,j} f(i-j) .
\end{align}
Expressing Sum \ref{eq:dif-sum-coef} in terms of $\delta$ along the same lines as we did in arriving at Equation \ref{eq:dif-sum-single} yields
\begin{align}\label{eq:dif-sum-coef-simp-delta}
\sum_{i=0}^{N} \, & \, \sum_{j=0}^{N} a_{i,j} f(i-j) \nonumber \\
  & = \sum_{k=-N}^{N} a_{k,k} f(0) \nonumber \\
  &+  \sum_{\delta=1}^N \sum_{k=0}^{N-\delta} \Big[ a_{k+\delta,k} f(\delta) +  a_{k,k+\delta} f(-\delta) \Big] .
\end{align}
The details of the math are not important but the student should verify that when $a_{i,j} = 1$ for all $i$ and $j$, the $k$ sum gives Equation \ref{eq:delta-number-times} yielding the full sum given by Equation \ref{eq:dif-sum-single}, as expected.

\subsection{Continuum Limit of Finite Sums}\label{sec:sum-to-int}

We seek to evaluate the sum
\begin{align}\label{eq:sum}
\sum_{\ell=j}^{k} f(\ell \omega_0) = \sum_{\ell=j}^{k} f(\ell \omega_0) \Delta \ell ,
\end{align}
where we use the fact that $\Delta \ell = (\ell+1) - \ell = 1$, so the continuum limit is given by
\begin{align}\label{eq:sum-to-continuum}
\sum_{\ell=j}^{k} f(\ell \omega_0) \Delta \ell \rightarrow \lim_{N \rightarrow \infty} \sum_{\ell = jN}^{kN} f(\frac {\ell} {N} \omega_0) \frac {\Delta \ell} {N} .
\end{align}
Defining $x=\ell/N$, Equation \ref{eq:sum-to-continuum} becomes
\begin{align}\label{eq:sum-to-integral}
\lim_{N \rightarrow \infty} \sum_{\ell= jN}^{kN} f(\frac {\ell} {N} \omega_0) \frac {\Delta \ell} {N} \rightarrow \int_j^k f(\omega_0 x) \, dx.
\end{align}

For convenience, we can make the variable change $\omega = \omega_0x$.  Then, the continuum limit of \ref{eq:sum-to-integral} becomes
\begin{align}\label{eq:sum-to-integral2}
\sum_{\ell=j}^{k} f(\ell \omega_0) \xrightarrow{\mbox{continuum}} \int_{j\omega_0}^{k\omega_0} d\omega \, \frac {f(\omega)} {\omega_0} = \int_{\omega_j}^{\omega_k} d\omega \, \frac {f(\omega)} {\omega_0} .
\end{align}

\section{Finding the Coefficients}\label{sec:GettingA}

We evaluate Equation \ref{eq:2-exponential-possible?} in the special case where the frequencies are equally spaced, that is, $\omega_{k^\prime} - \omega_k = \omega_0 (k^\prime - k)$ -- true for free particles and the harmonic oscillator in one dimension.  Such a spectrum is amenable to re-indexing the sums in terms of the difference $\delta = k^\prime - k$ using the procedure described in Appendix \ref{sec:sum-of-diff}.  This yields for the first equality in Equation \ref{eq:2-exponential-possible?}
\begin{align}\label{eq:2-exponential-possible-re-index}
& \rho_{n , n} (t)  = \lim_{N \rightarrow \infty} \sum_{k=0}^{N} \, \sum_{k^\prime=0}^{N} A_{k,k^\prime}^{(n)} e^{i \omega_0 \delta t}
  = \lim_{N \rightarrow \infty} \Big( \sum_{k=0}^{N} A_{k,k}^{(n)} \nonumber \\
&+  \sum_{\delta=1}^N \sum_{k=0}^{N-\delta} \Big[ A_{k + \delta,k}^{(n)} e^{+i \omega_0 \delta t} +  A_{k,k+\delta}^{(n)} e^{-i \omega_0 \delta t} \Big] \Big).
\end{align}

Defining
\begin{align}\label{eq:DefineB}
B_\delta^{(n)} = \sum_{k=0}^{N-\delta} A_{k + \delta,k}^{(n)}
\end{align}
and using Equation \ref{eq:AkkPrime-Hermetion}, Equation \ref{eq:2-exponential-possible-re-index} takes the simplified form
\begin{align}\label{eq:3-exponential-possible-re-index}
 \rho_{n , n} (t)  & = \lim_{N \rightarrow \infty} \Big( B_0^{(n)} \nonumber \\
  & +  \sum_{\delta=1}^N \Big[ B_\delta^{(n)} e^{+i \omega_0 \delta t} +  B_\delta^{(n)^*} e^{-i \omega_0 \delta t} \Big] \Big).
\end{align}

Next, we express Equation \ref{eq:3-exponential-possible-re-index} in the more compact form
\begin{align}\label{eq:3-compact}
 \rho_{n , n} (t)  & = \lim_{N \rightarrow \infty} \sum_{\delta=-N}^{+N} B_\delta^{(n)} e^{+i \omega_0 \delta t} ,
\end{align}
where evidently $ B_{-\delta}^{(n)} = B_\delta^{(n)^*}$.  Equation \ref{eq:3-compact} is transformed into the continuous limit as described in Appendix \ref{sec:sum-to-int} with Equation \ref{eq:sum-to-integral2}, resulting in the continuum form of Equation \ref{eq:2-exponential-possible?}:
\begin{align}\label{eq:2-exponential-possible-continuum}
\int_{-\infty}^{+\infty} \frac {d \omega} {\omega_0} B^{(n)} (\omega/\omega_0) \exp(i \omega t ), \stackrel{?}{=} \theta(t) b_n \exp(-\alpha t) ,
\end{align}
where we have added the step function $\theta(t)$ to set $t=0$ at the start of decay.

Multiplying both sides of Equation \ref{eq:2-exponential-possible-continuum} by $\exp(i \omega^\prime t)$, integrating over time and solving for $B^{(n)}$ yields
\begin{align}\label{eq:2-exponential-possible-continuum-integrate}
B^{(n)} (\omega^\prime / \omega_0)  = \frac {\omega_0} {2 \pi} \cdot \frac { b_n } {i \omega^\prime + \alpha} .
\end{align}
These are the coefficients of the two-particle superposition that yields a pure state that decays exponentially, showing that exponentials are compatible with unitary time evolution.


\begin{thebibliography}{10}
\newcommand{\enquote}[1]{``#1''}
\expandafter\ifx\csname url\endcsname\relax
  \def\url#1{{#1}}\fi
\expandafter\ifx\csname urlprefix\endcsname\relax\def\urlprefix{}\fi

\bibitem{sakur10.01}
J.~J. Sakurai and J.~J. Napolitano, {\em Modern Quantum Mechanics (2nd
  Edition)\/} (Addison-Wesley, 2010).
\newline
  http://www.amazon.com/Modern-Quantum-Mechanics-2nd-Edition/dp/0805382917

\bibitem{sakur94.01}
J.~J. Sakurai, {\em Modern Qunatum Mechanics - Revised Edition\/} (Addison
  Wesley Longman, 1994).

\bibitem{huang63.01}
K.~Huang, {\em Statistical Mechanics\/} (John WileyWiley and Sons, 1963).

\bibitem{pathr01.11}
R.~K. Pathria and P.~D. Beale, {\em Statistical Mechanics\/} (Elsevier Press,
  2011).

\bibitem{gottf01.74}
K.~Gottfried, {\em Quantum Mechanics\/}, Vol.~I (W. A. Benjamin, Inc., 1974).

\bibitem{patte62.01}
J.~D. Patterson, \enquote{Density Matrix Representations,} Am. J. Phys. {\bf
  30}, 894 (1962).

\bibitem{sabba07.01}
J.~Sabbaghzadeh and A.~Dalafi, \enquote{The role of the density operator in the
  statistical descriptionof quantum systems,} Am. J. Phys. {\bf 75}, 1162
  (2007).

\bibitem{neuma27.01}
J.~von Neumann, \enquote{Wahrscheinlichkeitstheoretischer Aufbau der
  Quantenmechanik,} G\"{o}ttinger Nachrichten {\bf 1}, 17 (1927).

\bibitem{neuma18.01}
W.~N.~A. von Neumann, John~and and R.~T. Beyer, {\em Mathematical Foundations
  of Quantum Mechanics: New Edition\/} (Princeton Universty Press, 2018).

\bibitem{tolma38.01}
R.~C. Tolman, {\em The Principles of Statistical Mechanics\/} (Oxford
  University Press, 1938), reprinted in 1979 by dover.

\bibitem{fano57.01}
U.~Fano, \enquote{Description of States in Quantum Mechanics by Density Matrix
  and Operator Techniques,} Reviews of Modern Physics {\bf 29}, 74--93 (1957).

\bibitem{landa69.01}
E.~M. Landau, L. D. \&~Lifshitz, {\em Statistical Physics\/} (Pergamon Press,
  1969), second revised english edition edn.

\bibitem{band79.01}
W.~Band and J.~L. Park, \enquote{Quantum state determination: Quorum for a
  particle in one dimension,} Am. J. Phys. {\bf 47}, 188--191 (1979).

\bibitem{landa77.01}
W.~M. Landau, L. D. \&~Lifshitz, {\em Quantum Mechanics (Non-relativistic
  Theory)\/} (Pergamon Press, 1977), third edition edn.

\bibitem{feynm72.01}
R.~P. Feynman, {\em Statistical Mechanics: A Set Of Lectures\/} (Basic Books,
  1972), 1st edn.

\bibitem{landau27.01}
L.~Landau, \enquote{Das D\"{a}mpfungsproblem in der Wellenmechanik (The damping
  problem in wave mechanics),} Z. Physik {\bf 22}, 430--441 (1980).

\bibitem{lerne17.01}
L.~Lerner, \enquote{A demonstration of decoherence for beginners,} Am. J. Phys.
  {\bf 85}, 870 (2017).

\bibitem{peres74.01}
A.~Peres, \enquote{Quantum Measurements Are Reversible,} Am. J. Phys. {\bf 42},
  886 (1974).

\bibitem{klein06.01}
M.~Kleinmann, H.~Kampermann, T.~Meyer, and D.~Bruss, \enquote{{Physical
  purification of quantum states},} {PHYSICAL REVIEW A} {\bf 73} ({2006}).

\bibitem{Nielsen01.10}
I.~L.~C. Michael A.~Nielsen, {\em Quantum Computation and Quantum
  Information\/} (Cambridge University Pr., 2010).

\bibitem{kuzyk19.01}
M.~G. Kuzyk, \enquote{Quantum no-cloning theorem and entanglement,} Am. J.
  Phys. {\bf 87}, 325--327 (2019).

\bibitem{duck98.01}
I.~Duck and E.~C.~G. Sudarshan, {\em Pauli and the spin-statistics theorem\/}
  (World Scientific, 1998).

\bibitem{schro17.01}
D.~V. Schroeder, \enquote{Entanglement isn't just for spin,} Am. J. Phys. {\bf
  85}, 812 (2017).

\bibitem{blum12.01}
K.~Blum, {\em Density matrix theory and applications\/} (Springer, 2012), 3rd
  edn.

\bibitem{weinb14.01}
S.~Weinberg, \enquote{Quantum mechanics without state vectors,} Phys Rev A {\bf
  90}, 042\,102 (2014).

\bibitem{gottes00.01}
D.~Gottesman and H.-K. Lo, \enquote{From Quantum Cheating to Quantum Security,}
  Physcs Today {\bf 53}, 22 (2000).

\bibitem{Beau17.01}
M.~Beau, J.~Kiukas, I.~Egusquiza, and A.~del Campo, \enquote{Nonexponential
  Quantum Decay under Environmental Decoherence,} Physical Review Letters {\bf
  119} (2017).

\bibitem{maxwe72.01}
N.~Maxwell, \enquote{A New Look at the Quantum Mechanical Problem of
  Measurement,} Am. J. Phys. {\bf 40}, 1431 (1972).

\bibitem{Popescu06.01}
S.~Popescu, A.~J. Short, and A.~Winter, \enquote{Entanglement and the
  foundations of statistical mechanics,} Nature Physics {\bf 2}, 754--758
  (2006).

\end{thebibliography}

\end{document}